\documentclass[prb,showpacs,amsmath,amssymb,floatfix,superscriptaddress,reprint]{revtex4-1}
\usepackage{graphicx}
\usepackage{mathtools}
\usepackage[makeroom]{cancel}
\usepackage{amssymb}
\usepackage{dcolumn}
\usepackage{fullpage}
\usepackage{amsmath}
\usepackage{multirow}
\usepackage{nicefrac}
\usepackage{physics}
\usepackage{dcolumn}
\usepackage{bm}
\usepackage{color}
\usepackage{upgreek}
\usepackage{hyperref}
\hypersetup
{
	colorlinks=true,
	allcolors=blue,
}

\newcommand{\irow}[1]   
{
  \begin{bmatrix} #1 \end{bmatrix}^\text{T}
}

\begin{document}

\title{Emergence of Intra-Particle Entanglement and \\ Time-Varying Violation of Bell's Inequality in Dirac Matter}

\author{Bruna Gabrielly de Moraes}
\affiliation{Catalan Institute of Nanoscience and Nanotechnology (ICN2), CSIC and BIST,
Campus UAB, Bellaterra, 08193 Barcelona, Spain}
\affiliation{Department of Physics, Universitat Aut\'{o}noma de Barcelona, Campus UAB,
Bellaterra, 08193 Barcelona, Spain}
\author{Aron W. Cummings}
\affiliation{Catalan Institute of Nanoscience and Nanotechnology (ICN2), CSIC and BIST,
Campus UAB, Bellaterra, 08193 Barcelona, Spain}
\author{Stephan Roche}
\email{stephan.roche@icn2.cat}
\affiliation{Catalan Institute of Nanoscience and Nanotechnology (ICN2), CSIC and BIST,
Campus UAB, Bellaterra, 08193 Barcelona, Spain}
\affiliation{ICREA--Instituci\'{o} Catalana de Recerca i Estudis Avan\c{c}ats, 08010 Barcelona, Spain}

\begin{abstract}
We demonstrate the emergence and dynamics of intra-particle entanglement in massless Dirac fermions. This entanglement, generated by spin-orbit coupling, arises between the spin and sublattice pseudospin of electrons in graphene. The entanglement is a complex dynamic quantity but is generally large, independent of the initial state. Its time dependence implies a dynamical violation of a Bell inequality, while its magnitude indicates that large intra-particle entanglement is a general feature of graphene on a substrate. These features are also expected to impact entanglement between pairs of particles, and may be detectable in experiments that combine Cooper pair splitting with nonlocal measurements of spin-spin correlation in mesoscopic devices based on Dirac materials.
\end{abstract}

\maketitle

The importance of quantum entanglement was sparked by the seminal \textit{gedanken} experiment proposed in 1935 by Einstein, Podolsky and Rosen \cite{PhysRev.47.777}, who in an attempt to demonstrate the incompleteness of quantum mechanics actually inspired the study of nonlocal correlations between distant particles \cite{PhysRev.47.777, schrodinger_1935, Bell1964}. Beyond the fascinating fundamental debate, which found its resolution in the violation of Bell's inequality \cite{Aspect2002, RevModPhys.86.419}, quantum entanglement between internal degrees of freedom of particles has become key for understanding emerging topological phases in many-body physics \cite{Keimer2017, Weneaal3099}, as well as for controlling quantum states in the quest to realize quantum computing applications \cite{RevModPhys.81.865, RevModPhys.74.347, Acin2018}.

Spin is a quantity that can be entangled to create nonlocal quantum correlations, and there have been many proposals to generate spin-entangled electrons \cite{PhysRevLett.84.1035, PhysRevB.66.161320, Kawabata2001} or electron-hole pairs \cite{PhysRevLett.91.147901, PhysRevLett.94.186804, PhysRevB.77.115323} in solid state systems. These have stimulated a variety of experimental studies, including attempts to generate entanglement through the splitting of Cooper pairs \cite{Hofstetter2009, PhysRevLett.104.026801, PhysRevLett.107.136801, Schindele2012, Das2012, PhysRevLett.114.096602, Fulop2015}. To date however, the verification of nonlocal quantum correlations between fermions, via a violation of Bell's inequality, remains challenging \cite{Burkard_2007, PhysRevB.95.115134, PhysRevB.96.064520}. In contrast, by taking advantage of \textit{intra}-particle entanglement between the spatial and spin degrees of freedom of a single neutron, the violation of Bell's inequality has been demonstrated in neutron interferometry experiments \cite{Hasegawa2003, Klepp2014}. Finally, there are many other potential combinations of intra- and inter-particle entanglement, as well as proposals to swap between them \cite{Yonac2007, Adhikari_2010, Zhang2017, Kumari2019}.

Two-dimensional Dirac materials offer true novelty in this context. In addition to spin, electrons in Dirac materials such as graphene carry other degrees of freedom, including sublattice (aka pseudospin), valley, and layer (in bilayer graphene) \cite{Pesin2012, Xiao2012, Schaibley2016, Novoselovaac9439}, which opens new possibilities for intra-particle entanglement. Sharing the mathematical structure of spin, these degrees of freedom can become entangled with each other and with spin, such that manipulating one degree of freedom may result in an effect on another. For example, when graphene is on a substrate, Rashba spin-orbit coupling (SOC) leads to entanglement between the spin and pseudospin. Identified as a source of spin dephasing in pristine graphene \cite{vantuan2014np, vantuan2016scirep, Cummings2016}, this is a clear example of the effect of intra-particle entanglement on a measurable spin transport property. Finally, the eigenstates of bilayer graphene also exhibit strong sublattice-layer entanglement near the charge neutrality point \cite{PhysRevB.95.195145}. However, there are currently no studies of the formation and dynamics of intra-particle entanglement in 2D Dirac materials.

In this Communication we explore the origin, dynamics, and magnitude of intra-particle entanglement between the spin and pseudospin degrees of freedom of electrons propagating in graphene. We show that  Rashba SOC drives the generation and evolution of this entanglement, which emerges regardless of the initial quantum state. Large intra-particle entanglement is thus a general feature of graphene, opening the door for its indirect detection via inter-particle correlations, and future research on its use in alternative quantum information processing schemes.

\begin{figure}[t]
\centering
\includegraphics[width=\columnwidth]{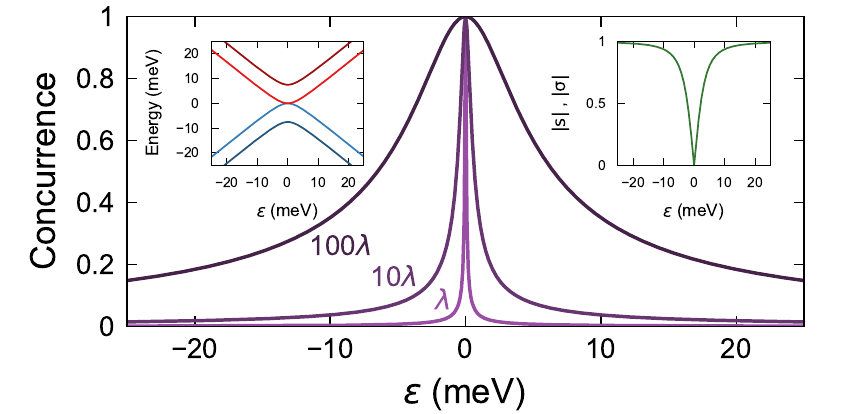}
\caption{Concurrence of the eigenstates of the graphene-Rashba system near the charge neutrality point for different strengths of Rashba SOC. The left inset shows a typical band structure and the right inset shows the magnitude of the spin and pseudospin of the eigenstates, both for $\lambda_\text{R} = 100\lambda = 3.75$ meV.}
\label{fig:fig1}
\end{figure}

\textit{Hamiltonian and entanglement of eigenstates}.\ We consider a continuum model of graphene with Rashba SOC, which is induced by a perpendicular electric field or a substrate. The Hamiltonian is
\begin{align}
\hat{\cal H} &= \hbar v_\text{F} \left( \tau \hat{\sigma}_x k_x + \hat{\sigma}_y k_y \right) \otimes \hat{s}_0 \nonumber \\
&+ \lambda_\text{R} \left( \tau \hat{\sigma}_x \otimes \hat{s}_y - \hat{\sigma}_y \otimes \hat{s}_x \right),
\label{eq:hamiltonian}
\end{align}
where $v_\text{F}$ is the Fermi velocity, $\tau = \pm 1$ is the valley index, $\hbar\bm{k}$ is the electron momentum, $\lambda_\text{R}$ is the Rashba SOC strength, and $\bm{\hat{s}}$ ($\bm{\hat{\sigma}}$) are the Pauli matrices for the spin (pseudospin) degree of freedom. The eigenenergies of $\hat{\cal H}$ are $\varepsilon_\pm^\text{e,h} = \nu \varepsilon_\pm$, where $\nu = \pm 1$ for electrons/holes, $\varepsilon_\pm = \sqrt{\varepsilon^2 + \lambda_\text{R}^2} \pm \lambda_\text{R}$, and $\varepsilon = \hbar v_\text{F} |\bm{k}|$. A typical bandstructure is shown in the left inset of Fig.\ \ref{fig:fig1}, with the conduction and valence bands split by $2\lambda_\text{R}$.

We limit ourselves to a single valley ($\tau = 1$) and express the Hamiltonian in the basis $\{ \ket{A\uparrow} , \ket{B\uparrow} , \ket{A\downarrow} , \ket{B\downarrow} \}$, with states of the form $\irow{ w_{A\uparrow} & w_{B\uparrow} & w_{A\downarrow} & w_{B\downarrow} }$, which describes the relative weights of the wave function on the A or B sublattice, with $\uparrow$ ($\downarrow$) denoting spin pointing along $+z$ ($-z$), perpendicular to the graphene plane.
Like the spin, the pseudospin can point in an arbitrary direction on the Bloch sphere, given the proper distribution of the wave function between the $A$ and $B$ sublattices. For example, the state $\frac{1}{\sqrt{2}} \irow{1&1&0&0} = \frac{1}{\sqrt{2}} \left( \ket{A}+\ket{B} \right) \otimes \ket{\uparrow}$ has pseudospin pointing along $+x$ and spin along $+z$, while $\frac{1}{2} \irow{1&-\text{i}&-1&\text{i}} = \frac{1}{\sqrt{2}} \left( \ket{A}-\text{i}\ket{B} \right) \otimes \frac{1}{\sqrt{2}} \left( \ket{\uparrow}-\ket{\downarrow} \right)$ has pseudospin along $-y$ and spin along $-x$. In this basis the eigenstates of the Hamiltonian are
\begin{equation}
\ket{\phi_\pm^\text{e,h}} = \frac{1}{\sqrt{N_\pm}} \irow{ \text{e}^{-\text{i}\theta} & \nu\gamma_\pm & \pm\text{i}\gamma_\pm & \pm\nu\text{i}\text{e}^{\text{i}\theta} },
\label{eq:eigstates}
\end{equation}
where $\theta = \arctan(k_y/k_x)$ is the direction of electron momentum, $\gamma_\pm = \varepsilon_\pm / \varepsilon$, and $N_\pm = 2(1+\gamma_\pm^2)$.

To quantify entanglement between the spin and pseudospin, we use the concurrence $C_\psi$ of a given state $\ket{\psi}$ \cite{PhysRevLett.80.2245}. The concurrence has a 1-to-1 relationship with the von Neumann entropy, equals 0 for completely separable states, and equals 1 for maximally entangled states. This entanglement measure was originally defined for mixed two-qubit systems, and has been extended to systems of many qubits \cite{Ma2011}. Here we study pure states of the form $\ket{\psi} = \irow{a&b&c&d}$, where the two qubits are the spin and pseudospin of the electron. In this case the concurrence is $C_\psi = 2|ad-bc|$.

Applying this definition to Eq.\ \eqref{eq:eigstates} gives $C_{\phi_\pm^\text{e,h}} = \lambda_\text{R} / \sqrt{\varepsilon^{2} + \lambda_\text{R}^{2}}$. The eigenstates of the graphene-Rashba system are thus maximally entangled near the charge neutrality point ($C_\phi \rightarrow 1$ as $\varepsilon \rightarrow 0$), and the entanglement decays as $1/\varepsilon$ at finite doping. This is shown in Fig.\ \ref{fig:fig1} for different values of $\lambda_\text{R} = n \lambda$, with $\lambda = 37.5$ $\upmu$eV typical of graphene on SiO$_2$ or hBN \cite{Zollner2019}, and $10\lambda$ or $100\lambda$ typical of graphene on high-SOC substrates \cite{garcia2018csr}. The total spin and pseudospin of the eigenstates, $|\bm{s}| = |\bm{\sigma}| = |\varepsilon| / \sqrt{\varepsilon^2 + \lambda_\text{R}^2}$, are shown in the right inset. Both disappear as $\varepsilon \rightarrow 0$, a result of their becoming maximally entangled.

\textit{Entanglement dynamics}.\ We now examine the entanglement between the spin and pseudospin of an arbitrary electronic state injected into graphene. This electron will be in an initial state $\ket{\psi} = \irow{a&b&c&d}$ and will evolve in time as $\ket{\psi(t)} = \hat{U}(t) \ket{\psi}$, where $\hat{U}(t) = \sum_j \exp(-\text{i} \varepsilon_j t / \hbar) \ket{\phi_j} \bra{\phi_j}$ is the time evolution operator, with $\varepsilon_j$ and $\ket{\phi_j}$ the eigenenergies and eigenstates of $\hat{\mathcal{H}}$, described above. The concurrence of this state will thus evolve in time as $C_\psi(t) = 2|a(t)d(t) - b(t)c(t)|$.

\begin{figure}[t]
\centering
\includegraphics[width=\columnwidth]{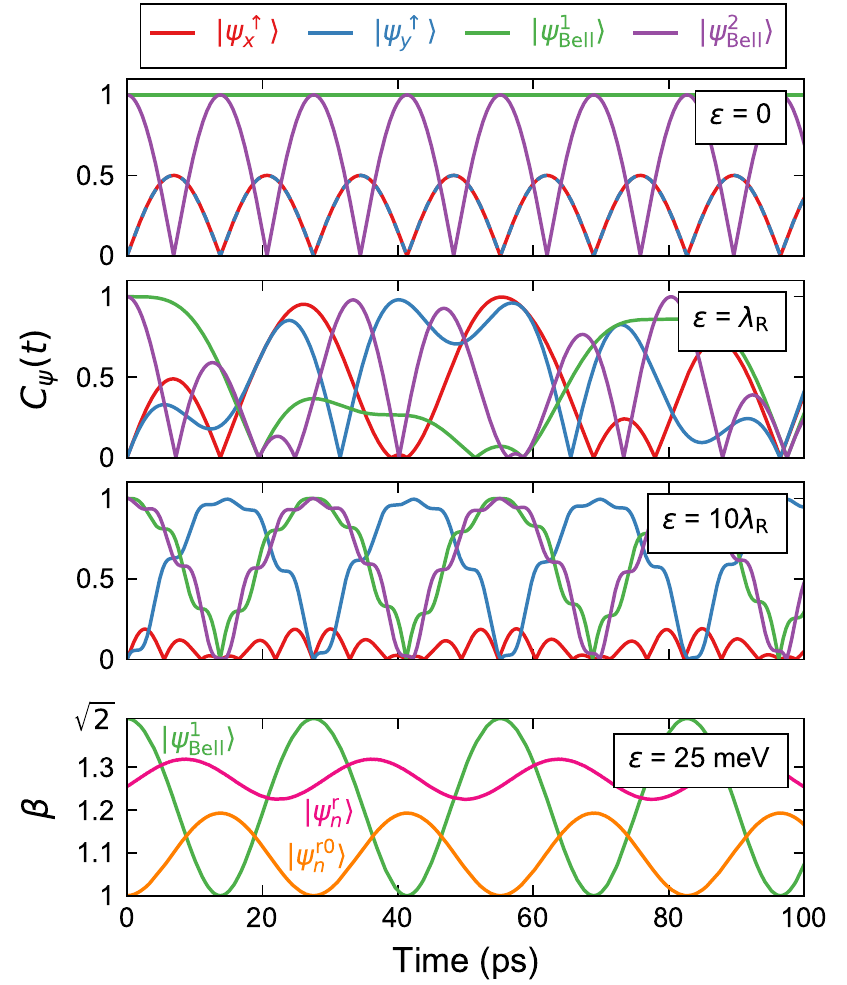}
\caption{Entanglement dynamics in the graphene-Rashba system. The top three panels show the concurrence dynamics of the initial states $\ket{\psi_{x,y}^\uparrow}$ and $\ket{\psi_\text{Bell}^{1,2}}$ at energies $\varepsilon = 0$, $\lambda_\text{R}$, and $10\lambda_\text{R}$, with $\lambda_\text{R} = 37.5$ $\upmu$eV. The bottom panel shows the time-dependent violation of the CHSH inequality for $\ket{\psi_\text{Bell}^{1}}$, one instance of $\ket{\psi_n^\text{r}}$, and one instance of $\ket{\psi_n^\text{r0}}$ at $\varepsilon = 25$ meV, also for $\lambda_\text{R} = 37.5$ $\upmu$eV.}
\label{fig:fig2}
\end{figure}

We first highlight the entanglement dynamics of some specific initial states.  Without loss of generality, we assume transport along the $x$-axis such that $\theta = 0$. Four states are considered : $\ket{\psi_x^\uparrow} = \frac{1}{\sqrt{2}} \irow{1&1&0&0}$, with pseudospin pointing along $+x$, parallel to the transport direction, and spin along $+z$; $\ket{\psi_y^\uparrow} = \frac{1}{\sqrt{2}} \irow{1&\text{i}&0&0}$, with pseudospin along $+y$, \textit{perpendicular} to the transport direction, and spin along $+z$; $\ket{\psi_\text{Bell}^1} = \frac{1}{\sqrt{2}} \irow{1&0&0&1}$, a Bell state with maximal spin-pseudospin entanglement; and $\ket{\psi_\text{Bell}^2} = \frac{1}{\sqrt{2}} \irow{0&1&1&0}$, another Bell state with maximal entanglement. The entanglement dynamics of these states are shown in Fig.\ \ref{fig:fig2} at three different energies $\varepsilon = 0$, $\lambda_\text{R}$, and $10\lambda_\text{R}$, with $\lambda_\text{R} = 37.5$ $\upmu$eV, depicted in the top three panels.

We first analyze the concurrence dynamics at the charge neutrality point, $\varepsilon = 0$. The Bell state $\ket{\psi_\text{Bell}^1}$ maintains perfect entanglement while $\ket{\psi_\text{Bell}^2}$ oscillates between perfect and zero entanglement. Interestingly, $\ket{\psi_{x,y}^\uparrow}$ both develop nonzero entanglement despite being initially separable, periodically reaching a maximum concurrence of 0.5. These behaviors can be understood by examining the eigenenergies and eigenstates of $\hat{\mathcal{H}}$ at $\varepsilon = 0$,
\begin{align}
\ket{\phi_+^\text{e,h}} &= \frac{1}{\sqrt{2}}\irow{ 0 & \nu & \text{i} & 0 } \text{, with } \varepsilon_+^\text{e,h} = \pm 2 \lambda_\text{R}, \nonumber \\
\ket{\phi_-^\text{e,h}} &= \frac{1}{\sqrt{2}}\irow{ 1 & 0 & 0 & -\text{i}\nu } \text{, with } \varepsilon_-^\text{e,h} = 0.
\label{eq:eig_cnp}
\end{align}
The eigenstates are maximally entangled, but their linear combinations can yield different behavior. The first Bell state can be written as $\ket{\psi_\text{Bell}^1} = \frac{1-\text{i}}{\sqrt{2}} \ket{\phi_-^\text{e}} + \frac{1+\text{i}}{\sqrt{2}} \ket{\phi_-^\text{h}}$. These eigenstates are degenerate ($\varepsilon_-^\text{e,h} = 0$) and $\ket{\psi_\text{Bell}^1}$ thus remains static with its initial $C_{\psi_\text{Bell}^1} = 1$. Meanwhile, the second Bell state can be written as $\ket{\psi_\text{Bell}^2} = \frac{1-\text{i}}{\sqrt{2}} \ket{\phi_+^\text{e}} + \frac{1+\text{i}}{\sqrt{2}} \ket{\phi_+^\text{h}}$. These eigenstates are separated in energy by $\varepsilon_+^\text{e} - \varepsilon_+^\text{h} = 4\lambda_\text{R}$, and $\ket{\psi_\text{Bell}^2}$ oscillates with $C_{\psi_\text{Bell}^2}(t) = |\cos(4\lambda_\text{R}t/\hbar)|$. The separable states can be written as $\ket{\psi_x^\uparrow} = \frac{1}{2}\left(\ket{\phi_+^\text{e}} - \ket{\phi_+^\text{h}}\right) + \frac{1}{2}\left(\ket{\phi_-^\text{e}} + \ket{\phi_-^\text{h}}\right)$ and $\ket{\psi_y^\uparrow} = \frac{\text{i}}{2}\left(\ket{\phi_+^\text{e}} - \ket{\phi_+^\text{h}}\right) + \frac{1}{2}\left(\ket{\phi_-^\text{e}} + \ket{\phi_-^\text{h}}\right)$. The first term contributes an oscillatory component while the second term is static. The static component has zero concurrence ($\ket{\phi_-^\text{e}} + \ket{\phi_-^\text{h}} \propto \irow{1&0&0&0}$), limiting the maximum concurrence that is reached, with $C_{\psi_{x,y}^\uparrow}(t) = \frac{1}{2} |\sin(4\lambda_\text{R}t/\hbar)|$.

Next we analyze the concurrence dynamics when approaching the high-energy limit, which is the most likely experimental situation (third row in Fig.\ \ref{fig:fig2}). Here the concurrence of both Bell states oscillates as $C_{\psi_\text{Bell}^{1,2}}(t) \approx |\cos(2 \lambda_\text{R} t / \hbar)|$. Meanwhile, the concurrence of $\ket{\psi_x^\uparrow}$ is small, approaching zero as $\varepsilon \rightarrow \infty$. In contrast, the concurrence of $\ket{\psi_y^\uparrow}$ is $C_{\psi_y^\uparrow}(t) \approx |\sin(2 \lambda_\text{R} t / \hbar)|$, and periodically reaches \textit{maximal} entanglement. Remarkably, in graphene with Rashba SOC, an electron that initially has no intra-particle entanglement can become maximally entangled, and has the same average entanglement as the maximally-entangled Bell states.

To understand the origin of this, we now look at the problem from the perspective of spin and pseudospin dynamics. In the graphene-Rashba system, the spin and pseudospin will precess around effective magnetic and pseudomagnetic fields. By examining the Hamiltonian in Eq.\ \eqref{eq:hamiltonian} and considering transport along $x$, these effective fields are \cite{vantuan2014np}
\begin{align}
\bm{B}_{\bm{s}}^{\text{eff}}(t) &= \lambda_\text{R} \left( -\langle\hat{\sigma}_y\rangle(t) \,,\, \langle\hat{\sigma}_x\rangle(t) \,,\, 0 \right), \nonumber \\
\bm{B}_{\bm{\sigma}}^{\text{eff}}(t) &= \lambda_\text{R} \left( \langle\hat{s}_y\rangle(t) \,,\, -\langle\hat{s}_x\rangle(t) \,,\, 0 \right) \\
&+ \varepsilon \left( 1\,,\,0\,,\,0 \right), \nonumber
\end{align}
where $\langle \hat{o} \rangle(t) \equiv \bra{\psi(t)} \hat{o} \ket{\psi(t)}$. In the presence of Rashba SOC, the spin and pseudospin precess around one another with a frequency $\omega_\text{R} = 2\lambda_\text{R}/\hbar$. The pseudospin also precesses around a component of $\bm{B}_{\bm{\sigma}}^{\text{eff}}$ that is parallel to the momentum and has magnitude $\varepsilon$. In the high-energy limit ($\varepsilon \gg \lambda_\text{R}$), this term dominates the pseudomagnetic field. Thus, when the initial pseudospin points along the momentum direction, as for $\ket{\psi_x^\uparrow}$, the pseudospin remains fixed along $x$ and only the spin precesses, in the $x$-$z$ plane. The spin and pseudospin thus remain well-defined and separable at all times, and the entanglement is negligible. On the other hand, when the initial pseudospin is \textit{perpendicular} to the momentum direction, as for $\ket{\psi_y^\uparrow}$, the pseudospin precesses rapidly in the $y$-$z$ plane with frequency $\omega_\varepsilon = 2 \varepsilon / \hbar$. This then enables mutual precession between spin and pseudospin driven by Rashba SOC, and the development of entanglement on the time scale $\omega_\text{R}^{-1}$. In this way, a separable state whose pseudospin is not parallel to the direction of transport will develop finite intra-particle entanglement.

At intermediate energies where neither $\varepsilon$ nor $\lambda_\text{R}$ are dominant, more complex dynamics can emerge. This can be seen in second row of Fig.\ \ref{fig:fig2}, where $\varepsilon = \lambda_\text{R}$. Here, multiple precession processes all coexist with similar weight and frequency, giving rise to more complex dynamics. We finally note that this behavior may be connected to the concept of quantum synchronization. Recent work has found that synchronization between pairs of spins certifies the presence of entanglement \cite{Roulet2018}, which resembles the situation at $\varepsilon = 0$, while the absence of synchronization does not preclude the development of entanglement, as we see in the high-energy limit.

\begin{figure}[t]
\centering
\includegraphics[width=\columnwidth]{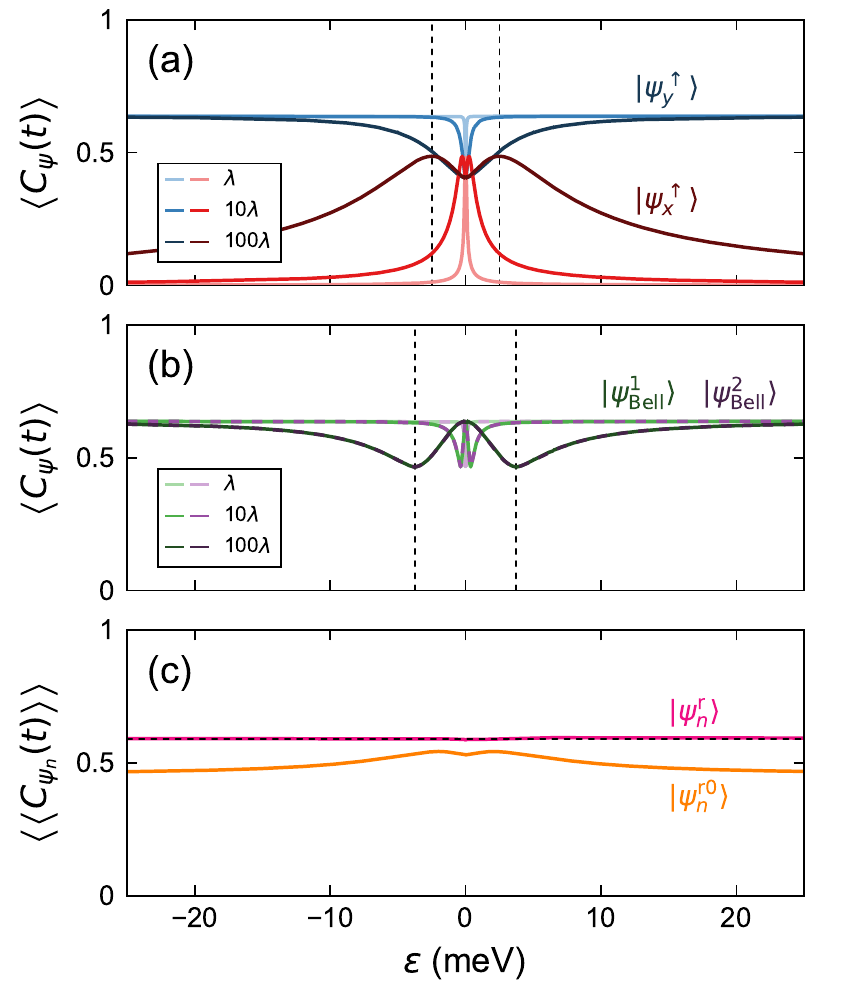}
\caption{Time-averaged concurrence as a function of energy for the initial states (a) $\ket{\psi_{x,y}^\uparrow}$ and (b) $\ket{\psi_\text{Bell}^{1,2}}$. For each state there are three curves, corresponding to Rashba SOC strengths of $\lambda_\text{R} = \lambda$, $10\lambda$, and $100\lambda$, with $\lambda = 37.5$ $\upmu$eV. Vertical dashed lines mark local maxima or minima in the concurrence for the case $\lambda_\text{R} = 100\lambda$. Panel (c) shows the time- and state-averaged concurrence, averaged over 1000 initial states $\ket{\psi_n^\text{r}}$ and $\ket{\psi_n^\text{r0}}$. The dashed line shows the average concurrence of the initial states $\ket{\psi_n^\text{r}}$ at $t=0$.}
\label{fig:fig3}
\end{figure}

\textit{Time-averaged entanglement}.\ Figure \ref{fig:fig2} shows that the entanglement can vary significantly with energy while exhibiting complex dynamics. To quantify the overall degree of entanglement of a given state at a given energy, we consider the time-averaged concurrence $\langle C_\psi(t) \rangle$. This is shown in Fig.\ \ref{fig:fig3}, with panel (a) corresponding to the initial states $\ket{\psi_{x,y}^\uparrow}$ and panel (b) to $\ket{\psi_\text{Bell}^{1,2}}$. For each state there are three curves, corresponding to $\lambda_\text{R} = \lambda$, $10\lambda$, and $100\lambda$.

At low energies, the average concurrences of $\ket{\psi_{x,y}^\uparrow}$ exhibit a local minimum, increasing away from $\varepsilon = 0$ as more complex dynamics come into play, as seen by comparing the top two panels of Fig.\ \ref{fig:fig2}. At higher energies the average concurrence of $\ket{\psi_{y}^\uparrow}$ continues to grow, approaching $\langle C_\psi(t) \rangle_\text{max} = \langle |\sin(\omega_\text{R}t)| \rangle = 2/\pi$. Meanwhile, for $\ket{\psi_{x}^\uparrow}$ the pseudomagnetic field dominates at higher energies, fixing the pseudospin and leading to a $1/\varepsilon$ decay of $\langle C_\psi(t) \rangle$. This crossover occurs at $|\varepsilon| = (2/3)\lambda_\text{R}$, marked by the vertical dashed lines in panel (a). In Fig.\ \ref{fig:fig3}(b), both Bell states have the same energy dependence (except exactly at $\varepsilon = 0$, see below). Around $\varepsilon = 0$, $\langle C_\psi(t) \rangle$ approaches $2/\pi$ and decays with increasing energy. This can be seen in the top two rows of Fig.\ \ref{fig:fig2}, where the complex spin-pseudospin dynamics at finite $\varepsilon$ lead to an average reduction of the concurrence from its maximal behavior. The minimal average concurrence of the Bell states is reached at $|\varepsilon| = \lambda_\text{R}$, shown by the vertical dashed lines. Above this energy, the Bell states converge back to $\langle C_\psi(t) \rangle_\text{max} = \langle |\cos(\omega_\text{R}t)| \rangle = 2/\pi$.

Here we note that $\ket{\psi_\text{Bell}^{1}}$ has a constant $C_\psi = 1$ at $\varepsilon = 0$ (top row of Fig.\ \ref{fig:fig2}). However, this occurs only exactly at $\varepsilon = 0$, and we have left this data point out of Fig.\ \ref{fig:fig3}. For any $\varepsilon \ne 0$, the spin-pseudospin dynamics limit the average concurrence to $\langle C_\psi(t) \rangle < 2/\pi$. We also note that the Rashba SOC strength has no impact on the magnitude of $\langle C_\psi(t) \rangle$; its only effect is to rescale the energy dependence, as shown by the different curves in Fig.\ \ref{fig:fig3}(a,b).

\textit{Entanglement of a random state}.\ We have studied four specific states and found they all exhibit dynamic spin-pseudospin entanglement. Particularly interesting is that the initially separable state $\ket{\psi_y^\uparrow}$ exhibits as much time-averaged intra-particle entanglement as the maximally-entangled Bell states. This suggests that in order to generate intra-particle entanglement in graphene, it is not necessary to prepare entangled states. Instead, injecting appropriate ``normal'' states is sufficient. However, injecting an initial state with well-defined spin and pseudospin is also challenging. The spin can be controlled to some degree with ferromagnetic contacts, but precise control of the pseudospin, determined by the weight and phase of the wave function on each sublattice, is more difficult.

Given these issues, it is useful to consider a simpler situation, where a bulk metal contact injects an electrical current into graphene. This current will consist of electrons with random initial states, which raises the question: what is the entanglement of an arbitrary state in graphene? This can be quantified by averaging over many random initial states $\ket{\psi_n}$, such that $\langle\langle C_{\psi_n}(t) \rangle\rangle \equiv \frac{1}{N} \sum_{n=1}^N \langle C_{\psi_n}(t) \rangle$ is the average intra-particle entanglement in graphene. We consider two sets of random initial states,
\begin{align}
\ket{\psi_n^\text{r}} &= \irow{a_n&b_n&c_n&d_n}, \nonumber \\
\ket{\psi_n^\text{r0}} &= \left\{ \cos(\theta_n^\text{p}/2)\ket{A} + \sin(\theta_n^\text{p}/2)\text{e}^{\text{i}\phi_n^\text{p}}\ket{B} \right\} \nonumber \\
&\otimes \left\{ \cos(\theta_n^\text{s}/2)\ket{\uparrow} + \sin(\theta_n^\text{s}/2)\text{e}^{\text{i}\phi_n^\text{s}}\ket{\downarrow} \right\}.
\end{align}
Each element of $\ket{\psi_n^\text{r}}$ is a random complex number chosen from the normal distribution. This generates states that are equivalent to the action of a random unitary matrix on some reference state, which are uniform over the four-dimensional Hilbert space \cite{Zyczkowski2001}, with average initial concurrence $\langle C_{\psi_n^\text{r}}(0) \rangle \approx 0.59$. Meanwhile, $\ket{\psi_n^\text{r0}}$ is a separable state, with $(\theta_n^\text{p} , \phi_n^\text{p})$ and $(\theta_n^\text{s} , \phi_n^\text{s})$ random spherical angles defining the orientations of the pseudospin and spin on the Bloch sphere. The set $\left\{ \ket{\psi_n^\text{r0}} \right\}$ thus represents all two-qubit states with zero initial entanglement.

The averages over these initial states are shown in Fig.\ \ref{fig:fig3}(c). For all possible initial states, the average concurrence is independent of energy and equal to its initial value, $\langle\langle C_{\psi_n^\text{r}}(t) \rangle\rangle = \langle C_{\psi_n^\text{r}}(0) \rangle \approx 0.59$, shown by the magenta and the dashed black lines. Meanwhile, the average over separable initial states has an energy dependence similar to $\langle C_{\psi_x^\uparrow}(t) \rangle$ in Fig.\ \ref{fig:fig3}(a), but with a slower decay. Importantly, $\langle\langle C_{\psi_n^\text{r0}}(t) \rangle\rangle$ remains large over the entire energy range, around 0.5. This indicates that large intra-particle entanglement is a general feature of graphene with spin-orbit coupling, even for states that are initially separable.

\textit{Time-dependent Bell inequality violation}.\ As shown above, intra-particle entanglement between spin and pseudospin is generally large in graphene with Rashba SOC. We have used the concurrence to quantify entanglement, but experimentally this is not directly measurable. Rather, entanglement is demonstrated through a violation of a Bell inequality \cite{Aspect2002, RevModPhys.86.419}. For pure states there is a direct connection between the concurrence and maximal violation of the Clauser-Horne-Shimony-Holt (CHSH) variant of Bell's inequality \cite{chsh1969}, given by $\beta = \sqrt{1+C_\psi^2}$ \cite{Verstraete2002}. Thus, because the concurrence of a particular electron is time-dependent, its violation of the CHSH inequality also varies in time. We show this explicitly in Fig.\ \ref{fig:fig2}(d), where we plot the time dependence of $\beta$ for $\ket{\psi_\text{Bell}^1}$, one instance of $\ket{\psi_n^\text{r}}$, and one instance of $\ket{\psi_n^\text{r0}}$ at $\varepsilon = 25$ meV. In all cases, $\beta$ oscillates with frequency $\omega_\text{R}$, highlighting the point that the Bell inequality violation is a dynamic and periodic quantity in the graphene-Rashba system.

\textit{Discussion and conclusions}.\ We have shown that intra-particle entanglement in graphene is a complex dynamic quantity, governed by the mutual precession of spin and pseudospin. For an arbitrary initial state the average entanglement is large, with a concurrence of $0.5$--$0.6$ that corresponds to a $10$--$16\%$ maximal violation of the CHSH inequality. Most importantly, spin-orbit coupling drives the generation of entanglement in electrons that are initially separable. This suggests that intra-particle entanglement in graphene may be robust to disorder and dephasing; after an entanglement-destroying interaction, spin and pseudospin dynamics will lead to re-generation of the intra-particle entanglement. The interplay between this entanglement re-generation and entanglement revival in a dephasing environment \cite{Leggio2015} may also yield unique behavior.

Going beyond the single-particle picture, these results have important implications for inter-particle entanglement between pairs of electrons in graphene. It has been shown that the dynamics of intra-particle entanglement can significantly impact inter-particle entanglement, including transfer from one type of entanglement to another \cite{Yonac2007}. Experimentally, this could be realized via the injection and splitting of spin-entangled Cooper pairs in graphene-based devices \cite{PhysRevLett.104.026801, PhysRevLett.114.096602}. The dynamical evolution of intra-particle entanglement would then be reflected in the inter-particle spin-spin correlation, and could be detectable in conventional nonlocal setups \cite{Hofstetter2009, PhysRevLett.104.026801, PhysRevLett.107.136801, Schindele2012, Das2012, PhysRevLett.114.096602, Fulop2015}, either using magnetic detectors \cite{Kawabata2001, PhysRevB.96.064520} or current correlation (noise) measurements \cite{Guido2000, Burkard_2007, Kindermann2009}, with further tuning of the dynamics via magnetic fields \cite{Nikitin2015}. If these dynamical aspects of the Bell inequality violation could be detected in such experiments, it would open a hitherto unexplored dimension concerning generation, detection, and manipulation of entanglement in Dirac and topological materials.

\clearpage

\begin{acknowledgments}
We thank L.E.F.\ Torres for fruitful comments. ICN2 is funded by the CERCA Programme/Generalitat de Catalunya, and is supported by the Severo Ochoa program from Spanish MINECO (Grant No.\ SEV-2017-0706). The authors acknowledge the European Union's Horizon 2020 research and innovation programme under grant agreement Nos.\ 785219 and 881603 (Graphene Flagship). B.G.\ de M.\ has received funding from the European Union's Horizon 2020 research and innovation programme under the Marie Sk\l{}odowska-Curie grant agreement No.\ 754558 (PREBIST).
\end{acknowledgments}

\bibliographystyle{apsrev4-1}
\bibliography{manuscript_bib}

\end{document}